\documentclass[intlimits,twoside,a4paper]{article}

\usepackage{amsmath,amssymb}
\usepackage{graphicx}

\usepackage[T2A]{fontenc}
\usepackage[cp1251]{inputenc}

\usepackage[eqsecnum]{cmpj2}

\articletype{Regular article}

\title[resistance on cobweb networks]%
{Exact asymptotic expansion for the resistance between the center node and a node on the cobweb network boundary
}

\author[Izmailian \& Kenna]{Nickolay Izmailian\refaddr{label1,label2} and
        Ralph Kenna\refaddr{label2}}
\addresses{
\addr{label1} Yerevan Physics Institute, Alikhanian Brothers 2, 375036 Yerevan, Armenia
\addr{label2} Applied Mathematics Research Center, Coventry University, Coventry CV1 5FB, England
}

\begin{document}

\maketitle

\begin{abstract}
We analyze the resistance between two notes in a cobweb network of resistors.
Based on an exact expression, we  derive the asymptotic expansions for the resistance between the center node and a node on the boundary of the $M \times N$  cobweb network with resistors $r$ and $s$ in the two spatial directions.
All coefficients in this expansion are expressed through analytical functions.%
\keywords Resistor network; Asymptotic expansion

\pacs 05.50.+q, 05.60.Cd, 02.30.Mv
\end{abstract}

\section{Introduction}
\label{Introduction}

The classic problem in electrical circuit theory, first studied by Kirchhoff in 1847, is the calculation of the resistance between two arbitrary nodes in a resistor network \cite{kirch}. Besides its long-standing importance in electric circuit theory, the computation of resistances is also connected to a wide range of problems as diverse as random walks \cite{resistor2,Lovasz1996}, first-passage processes \cite{Redner2001}, lattice Green's functions \cite{Katsura1971} and classical transport in disorder media \cite{resistor4,resistor5,resistor6}.

In 2004 Wu \cite{Wu2004} derived a closed-form expression for the two-point resistance in terms of the eigenvalues and eigenvectors of the Laplacian matrix associated with the network.
Quite recently, Izmailian, Kenna and Wu \cite{IKW2014} revisited the problem of two-point resistance and derived a new and simpler expression for the resistance between two arbitrary nodes for finite networks with resistors $r$ and $s$ in the two spatial directions. The new expression was then applied to the cobweb resistor network \cite{IKW2014}.

Essam and Wu \cite{EssamWu2009} used one of the results  \cite{Wu2004} to  derived the asymptotic expansion for the corner-to-corner resistance $R(r,s)$ on an $M\times N$ rectangular resistor network under free boundary conditions. This was extended by Izmailian and Huang  \cite{IHuang} to other boundary conditions. In recent decades the finite-size scaling and finite-size corrections in finite
critical systems and their boundary effects have attracted much attention  \cite{Blote,Cardy1,Ietal1,Ietal2,Ietal3,Ietal4,Ietal5,Ietal6,Ietal7,Ietal8,Ietal9,Ietal10,Ietal11,Ietal12,
izmailian2002,izmailian2002a,izmailian2003,izmailian2007,EssamWu2009,IHuang}. Of particular importance in such studies are exact results wherein the analysis can be carried out without numerical errors.

In this paper we derive the exact asymptotic expansion for the resistance between the central node and a node on the boundary of the cobweb network. We show that this expansion can be written in the form
\begin{eqnarray}
\frac{1}{s}R(r,s)&=& c(h)\, \ln{S}+c_0(h,\xi)
+\sum_{p=1}^{\infty} \frac{c_{2p}(h,\xi)}{S^{p}},
\label{RmnAsymptotic1}
\end{eqnarray}
%as
%\begin{eqnarray}
%\frac{1}{s}R(r,s)&=& c(h)\, \ln{S}+c_0(h,\xi)
%+\sum_{p=1}^{\infty} \frac{c_{2p}(h,\xi)}{S^{p}}
%\label{RmnAsymptotic}
%\end{eqnarray}
with $h = s/r$, $S = (M+1/2) N$  and $\xi = (M+1/2)/N$.
Note that, instead of the actual length $M$,  we have use effective length $(M+1/2)$.
{\sl All} coefficients in this expansion ($c(h), c_0(h,\xi), c_{2p}(h,\xi)$) are expressed through analytical functions. The computation of the asymptotic expansion of the resistance between two maximally separated nodes of a rectangular resistor network has been of interest for some time, as its value provides a lower bound to the resistance of compact percolation clusters in the Domany-Kinzel model of a directed percolation \cite{Domany1984}.

The organization of this paper is as follows.
Based on the exact expression for the resistance between arbitrary two nodes for finite cobweb resistor network obtained in \cite{IKW2014}
we express the resistance between the central node and node on the boundary of the network in terms of $G_{\alpha, \beta}(h,M,N)$ with $(\alpha, \beta)= (0, 1/2)$ (Sec. II).
We then extend the algorithm of Ivashkevich, Izmailian and Hu \cite{izmailian2002} to derive the exact asymptotic expansions for the resistance between the central node and a node on the boundary of the cobweb  resistor network and write down the expansion coefficients (Sec. III).
Finally, we  discuss our results in Sec. IV.

%%%%%%%%%%%%%%%%%%%%%%%%%%%%%%%%%%%%%%%%%%%%%%%%%%%%%%%%%%%%%%%%%%%%%%%%
\section{Two-dimensional resistor networks}
\label{Second part}
The resistor network can be regarded as a graph consisting from T nodes and let $R_{i,j}=R_{j,i}$ be the resistance of the resistor connecting nodes $i$ and $j$. Denote the nonzero eigenvalues and eigenvectors of the Laplacian of that network by $\lambda_i$ and ${\bf \Psi_i}=(\psi_{i1},\psi_{i2},...,\psi_{iT})$, respectively.  Then the resistance between nodes $i$ and $j$ can be written as \cite{Wu2004}
\begin{equation}
R_{i,j}=\sum_{k=2}^{T}\frac{\left|\psi_{ki}-\psi_{kj}\right|^2}{\lambda_k}.
\label{Rab}
\end{equation}
Let us consider the cobweb network.
The cobweb lattice ${\cal L}_{\rm cob}$ is an $M \times N$ rectangular lattice with periodic boundary conditions in one direction and nodes on one of the two boundaries in the other direction connected to an external common node.
Therefore there is a total of $M N+1$ nodes.
The example of an $M=3, N=8$ cobweb with resistors $s$ and $r$ in the two directions is shown in Fig. 1.
Topologically ${\cal L}_{\rm cob}$ is of the form of a wheel consisting of $N$ spokes and $M$ concentric circles.
We use the term Dirichlet-Neumann  to describe the boundary conditions along the innermost apex and outermost arc.
There has been considerable recent interest in studying the resistance in a cobweb network (see for example Refs. \cite{IKW2014,ETW2014,Tan2013}).

The closed-form expression for the resistance $R({\bf r}_1,{\bf r}_2)$ between two arbitrary nodes ${\bf r}_1=(x_1, y_1)$ and ${\bf r}_2=(x_2, y_2)$ for the cobweb network was obtained in \cite{IKW2014}.
In what follows, we will show that the resistance $R(O,A)$ between the central node $O=(0,0)$ and node on the boundary of the network $A=(x,M)$ can be expressed in terms
of $G_{0, 1/2}({\cal M},{\cal N})$ only,
\begin{eqnarray}
R^{\,\rm cob}(0,A) &=&-\frac{s}{2}+\frac{\sqrt{s r}}{4 S} G_{0,1/2}(2M+1,N),
\label{2dRfin}
%\\
%R^{\,\rm fan}(0,A) &=&-\frac{s}{2}+\frac{\sqrt{s r}}{S} G_{0,1/2}(x,2M+1,2N)
%\label{2dRfin1}
\end{eqnarray}
where $S = (M+1/2) N$ and $G_{\alpha, \beta}({\cal M},{\cal N})$ is given by
\begin{equation}
G_{\alpha, \beta}({\cal M},{\cal N}) =
{\cal M}\;{\tt~Re}\sum_{n=0}^{{\cal N}-1}f\left(\pi\frac{n+\alpha}{{\cal N}}\right)\;
{\rm coth}\left[{\cal M}\,\omega\left(\pi\frac{n+\alpha}{{\cal N}}\right)+i \pi
\beta\right]\label{Gab},
\end{equation}
for $(\alpha, \beta) \ne (0,0)$. The function $\omega(y)$ is given by:
\begin{equation}
\omega(y) = {\rm arcsinh}\sqrt{h} \sin y \label{omega}
\end{equation}
and the function $f(y)$ is given by
\begin{eqnarray}
f(y)&=&\frac{\sqrt{1+h\sin^2 y}}{\sin y}.
 \label{fxcyl}
\end{eqnarray}
%for cobweb network and given by
%\begin{eqnarray}
%f(y,x) \equiv f^{fan}(y,x)&=&\frac{\cos^2 \left(y(2x+1)\right)\sqrt{1+h\sin^2 y}}{\sin %y}
% \label{fxcyl1}
%\end{eqnarray}
%for fan network.

%%%%%%%%%%%%%%%%%%%%%%%%%%%%%%%%%%%%%%%%%%%%%%%%%%%%%%%%%
%%%%%%%%%%%%%%%%%%%%%%%%%%%%%%%%%%%%%%%%%%%%%%%%%%%%%%%%%%
\subsection{Cobweb network }
\begin{figure}[tbp]
  \begin{center}
  \includegraphics[width=0.8\textwidth]{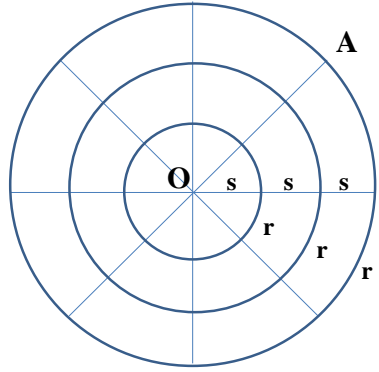}
  \end{center}
\vspace{-3cm}  \caption{An $M \times N$  cobweb network with $M=3$ and $N=8$.
  Bonds in the radial and circular directions comprise resistors $s$ and $r$.
  The center point is denoted by $O$, $A$ denotes any point on the boundary of the cobweb network.}
  \label{fig1}
  \end{figure}

 The resistance between the central node $O=(0, 0)$ and other node  $C=(x,y)$ of the cobweb network  is given by (see second line of Eq. (33) of Ref. \cite{IKW2014})
\begin{eqnarray}
R^{\rm{cob}}(O,C)
&=&
\frac{2s}{N(2M+1)} \sum_{m=0}^{M-1} \sum_{n=0}^{N-1}
\frac {\sin^2 ( 2y\varphi _m )}
{h(1-\cos2 \theta_n )+(1-\cos2 \varphi_m )}, \quad y = 1,2,...,M,
\label{ROAy}
\end{eqnarray}
where $h=s/r$ and
\begin{equation}
\theta_n=\frac{\pi n}{N} \qquad \qquad \varphi_m=\frac{\pi(m+1/2)}{2M+1} .
\label{thetavarphi}
\end{equation}
Note that the result (\ref{ROAy}) is independent of the position $x$ as it should be.

In the special case of the resistance between the center $O$ and a
point $A=\{x,N\}$ on the outer boundary of the cobweb network, we use $y=M$ and obtain from (\ref{ROAy})
\begin{equation}
R^{\rm cob}(O, A)=\frac{s}{N(2M+1)} \sum_{m=0}^{M-1} \sum_{n=0}^{N-1}
\frac {\cos^2 \varphi_m}
{\sin^2 \varphi_m+h\sin^2 \theta_n},
\label{ROAfinal}
\end{equation}
where use has been made of the identity
\begin{equation}
\sin ( 2\,M\,\varphi_m) = (-1)^{m} \cos \varphi_m, \nonumber
\end{equation}
 which is a consequence of the fact  $2M \varphi_m + \varphi_m  = \big( m+\frac 1 2 \big) \pi$.

Eq. (\ref{ROAfinal}) can be transformed as
\begin{eqnarray}
R^{\rm cob}(O, A)&=&\frac{s}{N(2M+1)} \sum_{m=0}^{M-1} \sum_{n=0}^{N-1}\left[-1+
\frac {1+h\sin^2 \theta_n}
{\sin^2 \varphi_m+h\sin^2 \theta_n}\right],
\label{ROAfinal1}\\
&=&-\frac{s M}{2M+1} +\frac{s}{N(2M+1)} \sum_{m=0}^{M-1} \sum_{n=0}^{N-1}
\frac {1+h\sin^2 \theta_n}
{\sin^2 \varphi_m+h\sin^2 \theta_n}.
\label{ROAfinal2}
\end{eqnarray}
We can extend the summation over  $m$ in  Eq. (\ref{ROAfinal2}) from $M-1$ up to $2M$ and obtain the expression
\begin{eqnarray}
R^{\rm cob}(O, A)
&=&-\frac{s}{2} +\frac{s}{2N(2M+1)} \sum_{m=0}^{2M} \sum_{n=0}^{N-1}
\frac {1+h\sin^2 \theta_n}
{\sin^2 \varphi_m+h\sin^2 \theta_n}.
\label{ROAfinal21}
\end{eqnarray}
The sum over $m$ in the Eq. (\ref{ROAfinal21}) can be carried out using the identity \cite{IHuang}
\begin{equation}
\sum_{m=0}^{{\cal M}-1} \Big[{h \sin^2 \theta_n +\sin^2
\frac{\pi (m+\frac{1}{2})}{{\cal M}}}\Big]^{-1}=2 {\cal M}\, \frac{{\rm
coth}\left[{\cal M}\,\omega(\theta_n)+i \pi/2\right]}{{\rm \sinh
2\omega(\theta_n)}} , \label{a5}
\end{equation}
with ${\cal M}=2M+1$ and $\omega(x)$  given by Eq. (\ref{omega}). It is easy
to see that
\begin{equation}
1+h
\sin^2{\theta_n}={\rm \cosh^2 \omega}(\theta_n) \label{2omega} .
\end{equation}

Plugging Eqs.  (\ref{a5}) and (\ref{2omega}) back in Eq.
(\ref{ROAfinal21}) we obtain that $R^{\rm cob}(O, A)$ can be written in the  form
\begin{eqnarray}
R^{\rm cob}(O, A)&=&-\frac{s}{2} +\frac{s}{2N}  \sum_{n=0}^{N-1}
\frac{{\rm
coth}\left[(2M+1)\,\omega(\theta_n)+i \pi/2\right]}{{\rm \tanh
\omega(\theta_n)}}.
\label{ROAfinal3}
\end{eqnarray}
Using identity
\begin{equation}
{\rm \tanh \omega}(\theta_n)=\frac{\sqrt{h}\,\sin{\theta_n}}{\sqrt{1+h
\sin^2{\theta_n}}},  \label{2omega1}
\end{equation}
 Eq. (\ref{ROAfinal3}) can be finally written in the form given by Eq. (\ref{2dRfin}).

%%%%%%%%%%%%%%%%%%%%%%%%%%%%%%%%%%%%%%%%%%%%%%%%%%%%%%%%%
%%%%%%%%%%%%%%%%%%%%%%%%%%%%%%%%%%%%%%%%%%%%%%%%%%%%%%%%%%

%%%%%%%%%%%%%%%%%%%%%%%%%%%%%%%%%%%%%%%%%%%%%%%%%%%%%%%%%
%%%%%%%%%%%%%%%%%%%%%%%%%%%%%%%%%%%%%%%%%%%%%%%%%%%%%%%%%%
\section{Asymptotic expansion}
In Sec. II we have shown that the  resistance between the central node and node on the boundary of the cobweb  resistor network can be expressed in terms of the function  $G_{0,1/2}(x,{\cal M},{\cal N})$  only, (see Eqs. (\ref{2dRfin})).
Using the method proposed in Ref. \cite{izmailian2002}, Izmailian and Huang \cite{IHuang}  derived the asymptotic expansion of $G_{\alpha,\beta}({\cal M},{\cal N})$ in terms of the so-called Kronecker double series \cite{Weil}, which are directly related to elliptic $\theta$ functions.
We next need the asymptotic expansion of  $G_{0,1/2}({\cal M},{\cal N})$, which can be found in Appendix \ref{Asymptotic}.

After reaching this point, one can easily write down all the terms of the exact asymptotic expansion for the resistance between the central node and a node on the boundary of the cobweb network ($R^{cob}(O,A)$) using Eqs. (\ref{2dRfin}) and (\ref{ExpansionGfin10}). We have found that the exact asymptotic expansion of the $R^{cob}(O,A)$  can be written as Eq. (\ref{RmnAsymptotic1}).

\subsection{Asymptotic expansion for the resistance between the central node and node on the boundary of the cobweb network}

For the cobweb network we obtain
\begin{eqnarray}
\frac{1}{s}R^{cob}(O,A) &=&\frac{1
}{2\pi\sqrt{h}}\left[\ln{S}+2\ln{\frac{8}{\pi}}+2
C_E-1-\ln{\xi(1+h)} +2\sqrt{h}\,{\rm
arctan}\sqrt{h}-4\ln{\theta_2(2i\sqrt{h}\,\xi)}\right]
\nonumber\\
&-& \frac{1}{2\pi\sqrt{h}} \sum_{p=1}^{\infty}\left(\frac{\pi^2
\xi}{S}\right)^{p}\frac{\Omega_{2p}}{p(2p)!} \, {\rm
K}_{2p}^{0,1/2}(2i\sqrt{h}\,\xi).
\label{ExpansionGfinfree}
\end{eqnarray}
Thus, the coefficients $c_{2p}(h,\xi)$ ($p$=1,2,..) in the expansion (\ref{RmnAsymptotic1}) are explicitly given by
\begin{equation}
c_{2p}(h,\xi)=-\frac{\pi^{2p-1} \xi^p}{2p(2p)!\sqrt{h}}
\Omega_{2p} \, {\rm K}_{2p}^{0,1/2}(2i\sqrt{h}\,\xi) \label{c2p}
\end{equation}
where the differential operators $\Omega_{2p}$ are given by Eq. (\ref{Omega2p}) and $K_{2p}^{0,1/2}(2i\sqrt{h}\xi)$ is Kronecker's double
series which can all be expressed in terms of the elliptic $\theta_k(2i\sqrt{h}\xi)$ ($k = 2, 3, 4$) functions only.

Here we list the first few coefficients in the expansion given by Eq. (\ref{RmnAsymptotic1}):
\begin{eqnarray}
c(h)&=&\frac{1}{2\pi\sqrt{h}}\label{cfree},\\
c_0(h,\xi)&=&\frac{1}{2\pi\sqrt{h}}\left(2\ln{\frac{8}{\pi}}+2
C_E-1-\ln{\xi(1+h)} +2\sqrt{h}\,{\rm
arctan}\sqrt{h}-4\ln{\theta_2}\right)
\label{c0free},\\
c_2(h,\xi)&=& \frac{\pi\tau_0}{288 h}\left((1+3 h)(\theta_3^4+\theta_4^4)+2
\tau_0 (1+h)\left(\pi \theta_3^4\theta_4^4+2(\theta_3^4+\theta_4^4)\frac{\partial}{\partial\tau_0}\ln
{\theta_2}\right)\right), \nonumber\\
~&\vdots&~ . \nonumber
\end{eqnarray}
To simplify the notation we have use the short hand
\begin{equation}
\theta_k=\theta_k(i \tau_0), \qquad k = 2, 3, 4,
\label{shorthand2}
\end{equation}
where $\tau_0=2\xi\sqrt{h}$.

We have also used the following relations between derivatives of the elliptic functions:
$$
\frac{\partial}{\partial \tau_0}\ln{{\theta}_3} =
\frac{\pi}{4}{\theta}_4^4+\frac{\partial}{\partial
\tau_0}\ln{{\theta}_2} \qquad \mbox{and} \qquad
\frac{\partial}{\partial \tau_0}\ln{{\theta}_4} =
\frac{\pi}{4}{\theta}_3^4+\frac{\partial}{\partial
\tau_0}\ln{{\theta}_2}.
$$
Note that elliptic functions ${\theta}_2, {\theta}_3, {\theta}_4$
can be expressed through the complete elliptic integral of the
first kind $K=K(k)$ and second kind $E=E(k)$ as
\begin{equation}
{\theta}_2=\sqrt{\frac{2 k K(k)}{\pi}}, \qquad
{\theta}_3=\sqrt{\frac{2 K(k)}{\pi}}, \qquad
{\theta}_4=\sqrt{\frac{2 k' K(k)}{\pi}} \label{K28}
\end{equation}
where
\begin{eqnarray}
K(k)&=& \int_0^{\pi/2}\frac{{\rm d}x}{\sqrt{1-k^2\sin^2{x}}},\label{EllipticIntK}\\
 E(k) &=& \int_0^{\pi/2}\sqrt{1-k^2\sin^2{x}}~\!{\rm d}x.
\label{EllipticIntE}
\end{eqnarray}

With the help of the identities
$$
\frac{\partial}{\partial \tau_0}\ln{\theta_2} =
-\frac{1}{2}{\theta}_3^2 E, \qquad \mbox{and} \qquad
\frac{\partial E}{\partial \tau_0}
=\frac{\pi^2}{4}{\theta}_3^2{\theta}_4^4-\frac{\pi}{2}{\theta}_4^4
E
$$
one can express all derivatives of the elliptic functions in terms
of the elliptic functions ${\theta}_2, {\theta}_3, {\theta}_4$ and
the complete elliptic integral of the second kind $E=E(k)$.

Thus we have obtained the explicit analytic formulas for all
corrections-to scaling terms $ c_{2p}(h,\xi)$ in the form of
elliptic functions. For the case $\xi=1$ and $h=1$ we have following results:
\begin{equation}
\frac{1}{s}R^{\rm cob}(O,A)=\frac{1}{\pi}\ln{N}+c_0+\frac{c_2}{N^2}+...
\label{wu}
\end{equation}
with $c_0=0.9286495235004523...$ and
$c_2=0.3572873939981...$.

\section{Discussion}
\label{last part}

In the present paper, we study the two-point resistor problem on
 the cobweb network. Using the exact expression for
the resistance between two arbitrary  nodes for finite cobweb network obtained in \cite{IKW2014} and the IIH's algorithm \cite{izmailian2002}, we derive the exact asymptotic expansion of the resistance between the central node and node on the boundary of the cobweb resistor networks. All coefficients in this
expansion are expressed through analytical functions.

\section{Acknowledgments}
\label{Acknowledgments}
The work was supported by a Marie Curie IIF (Project no. 300206-RAVEN)
and IRSES (Projects no. 295302-SPIDER and 612707-DIONICOS) within 7th European Community Framework
Programme and by the grant of the Science Committee of the Ministry of Science and
Education of the Republic of Armenia under contract 13-1C080.

\appendix

\section{Asymptotic expansion of $G_{0,1/2}(2M+1,N)$}
\label{Asymptotic}
The asymptotic expansion of $G_{\alpha,\beta}({\cal M},{\cal
N})$ for $(\alpha,\beta) \ne (0,0)$ has been obtained in Ref. \cite{IHuang}. Here we will reproduce the result of the paper Ref. \cite{IHuang} for the case $(\alpha,\beta) = (0,1/2)$, ${\cal M}=2M+1$ and ${\cal
N}=N$.
After little algebra the asymptotic expansion of $G_{0,1/2}(2M+1,N)$ can be written as
\begin{eqnarray}
G_{0,1/2}(2M+1,N)
&=&\frac{4 S}{\pi}\left[\frac{1}{2}\ln{\frac{S}{\xi}}+ C_E + \ln{\frac{8}{\pi}}-\frac{1}{2}\ln(1+h)+\sqrt{h}\,{\rm
arctan}\sqrt{h}-2
\ln{\left|\theta_{2}(2 i \xi \sqrt{h})\right|} \right]
\nonumber \\
&-&2 \pi \xi \sum_{p=1}^{\infty}\left(\frac{\pi^2 \xi}{
S}\right)^{p-1}\frac{\Omega_{2p}}{p(2p)!} {\tt Re}\; {\rm
K}_{2p}^{0,1/2}(2 i \xi \sqrt{h}) \label{ExpansionGfin10}
\end{eqnarray}
where $S=(M+1/2)N$, $\xi=(M+1/2)/N$, $C_E$ is the Euler constant, $\theta_2(\tau)$ is elliptic theta function and  $K_{2p+2}^{0,1/2}(\tau)$ is Kronecker's double series \cite{Weil}.

The differential operators $\Omega_{2p}$ that have appeared here
can be expressed via coefficients
$\omega_{2p}=\varepsilon_{2p}+\lambda_{2p}
\frac{\partial}{\partial\lambda}$ as
\begin{eqnarray}
{\Omega}_{2}&=&\omega_2\nonumber\\
{\Omega}_{4}&=&\omega_4+3\omega_2^2\, \label{Omega2p}\\ &\vdots&
\nonumber
\end{eqnarray}
where $\lambda_{2p}$ and $\kappa_{2p}$ are the coefficients in the Taylor expansion of  $\omega(y)$, given by Eq. (\ref{omega}) and $f(y)$ given by Eq. (\ref{fxcyl}), respectively
\begin{equation}
\omega(y)=y\left(\lambda+\sum_{p=1}^{\infty}
\frac{\lambda_{2p}}{(2p)!}\;y^{2p}\right) \label{omegaTaylor}
\end{equation}
with $\lambda=\sqrt{h}$, $\lambda_2=-\frac{1}{3}\sqrt{h}(1+h)$,
$\lambda_4=\frac{1}{5}\sqrt{h}(1+10 h+ 9h^2)$, etc,
and
\begin{equation}
f(y) = \frac{1}{y}\left[1+\sum_{p=1}^{\infty}\frac
{\kappa_{2p}}{(2p)!}y^{2p} \right],\label{omegaTaylor1}
\end{equation}
with $\kappa_{2}=-\frac{1}{3}-h$, $\kappa_{4}=-\frac{7}{15}+2h+3h^2$,
etc. Note that function $f(y)$ can be represented as
\begin{equation}
f(y)=\frac{1}{y}\;\exp\left\{
{\sum_{p=1}^{\infty}\frac{\varepsilon_{2p}}{(2p)!}}y^{2p} \right\},
\label{owega''}
\end{equation}
and the coefficients $\varepsilon_{2p}$ and $\kappa_{2p}$ are
related to each other through relation between moments and cumulants
\begin{eqnarray}
\kappa_{2}&=&\varepsilon_2\nonumber\\
\kappa_{4}&=&\varepsilon_4+3\varepsilon_2^2\,
\nonumber\\
&\vdots&  \nonumber
\end{eqnarray}

The Kronecker's double series $K_{2p}^{0,1/2}(\tau)$ can all be expressed in terms of the elliptic $\theta(\tau)$ functions only. Equations for $K_{2p}^{0,1/2}(\tau)$ with $p = 2, 3, 4, 5$
and other useful relations for elliptic $\theta$-functions and Kronecker's double series can be found in Refs. \cite{izmailian2002,izmailian2002a,izmailian2003,izmailian2007}.

%\end{document}

%% Type in your references using {thebibliography} environment
%% or create them from your bibtex database using cmpj.bst style (experimental).

%\bibliographystyle{cmpj}
%\bibliography{mybibdb}

%
%% If you have problems with typesetting in ukrainian uncomment lines below.
%
  \lastpage
  \end{document}